\documentclass[aps,prb,twocolumn,groupedaddress,floatfix]{revtex4-1}

\usepackage{graphicx}
\usepackage{dcolumn}
\usepackage{bm}

\begin{document}

\title{Manipulation of single electron spin in a GaAs quantum dot through the application of geometric phases: The Feynman disentangling technique}

\author{Sanjay Prabhakar,$^{1,2}$ James Raynolds,$^1$ Akira Inomata,$^3$ and Roderick Melnik$^{2,4}$}
\affiliation{
$^{1}$College of Nanoscale Science and Engineering, University at Albany,
State University of New York\\
$^2$M\,$^2$NeT Laboratory, Wilfrid Laurier University, Waterloo, ON, N2L 3C5 Canada\\
$^3$Department of Physics, University at Albany, State University of New York\\
$^4$BCAM, Bizkaia Technology Park, 48160 Derio, Spain
}

\date{\today}

\begin{abstract}
The spin of a single electron in an electrically defined
quantum dot in a $2$DEG can be manipulated by moving the quantum dot
adiabatically in a closed loop in the $2$D plane under the influence of
applied gate potentials. In this paper we present analytical expressions
and numerical simulations for the spin-flip probabilities during the
adiabatic evolution in the presence of the Rashba and Dresselhaus linear
spin-orbit interactions. We use the Feynman disentanglement technique to
determine the non-Abelian Berry phase and we find exact analytical
expressions for three special cases: (i) the pure Rashba spin-orbit coupling, (ii)
the pure Dresselhause linear spin-orbit coupling, and (iii) the mixture of the Rashba
and Dresselhaus spin-orbit couplings with equal strength. For a mixture
of the Rashba and Dresselhaus spin-orbit couplings with unequal strengths, we obtain simulation results by
solving numerically the Riccati equation originating from the
disentangling procedure. We find that the spin-flip probability in the
presence of the mixed spin-orbit couplings is generally larger than those for the
pure Rashba case and for the pure Dresselhaus case, and that the
complete spin-flip takes place only when the Rashba and Dresselhaus spin-orbit
couplings are mixed symmetrically.
\end{abstract}

\maketitle

\section{Introduction}
Geometric phases abound in physics and their study has attracted
considerable attention since the seminal work of Berry.\cite{berry84,wilczek_book}  
In recent years a number of researchers have shown their interest in the
geometric phases associated with single- and few-spin systems for
potential applications in the field of quantum computing and non-charge
based logic.\cite{loss98,jones00,hu00} 
One interesting proposal is the
notion that the spin of a single electron trapped in an
electrostatically defined $2$D quantum dot can be manipulated through the
application of gate potentials by moving the center of
mass of a quantum dot adiabatically in a closed loop and inducing a non-Abelian matrix Berry phase.\cite{jose06}
A recent work shows that the Berry phases can be changed
dramatically by the applications of gate potentials and may be detected in an
interference experiment.\cite{wang08}  

In the present paper, we study the non-Abelian unitary operator of the spin
states during the adiabatic motion of a single electron spin. The
non-Abelian nature here stems from the spin-orbit coupling of an
electron in two dimensions. The evolution operator which gives rise to
the Berry phase is not easy to evaluate as it contains non-commuting
operators. In 1951, Feynman\cite{feynman51} 
developed an operator calculus
for quantum electrodynamics, in which he devised a way to disentangle
the evolution operator involving non-commuting operators. In 1958, Popov\cite{popov58} 
applied the operator calculus, combined with group-theoretical considerations, to the spin rotation for a particle with a
magnetic moment in an external magnetic field to obtain exact transition
probabilities between the initial and final spin states. In a way
similar to Popov's we employ the Feynman technique to disentangle the
evolution operator for a quantum dot with the Rashba and Dresselhaus spin-orbit couplings and
derive analytical expressions for spin transition probabilities. In
particular, we obtain exact closed form expressions for three specific
cases: (i) the pure Rashba spin-orbit coupling\cite{bychkov84} 
(ii) the pure linear Dresselhaus spin-orbit coupling,\cite{dresselhaus55} 
and (iii) the symmetric combination of
the Rashba and Dresselhaus spin-orbit couplings. This approach provides us a
convenient numerical scheme for an arbitrary mixing of the two types of
spin-orbit couplings via a Riccati equation.\cite{riccati_book} 
An interesting result we
find is that the spin-flip probability for the case of an arbitrary
mixture of the Rashba and the Dresselhaus spin-orbit couplings is generally greater
than that for the case where either the Rashba or the Dresselhaus
interaction acts alone. Furthermore, we see that the complete spin
precession occurs only when the Rashba spin-orbit coupling and the Dresselhaus spin-orbit coupling are equal in strength.

The work of Berry teaches that if parameters contained in the
Hamiltonian of a quantal system are adiabatically carried around a
closed loop an extra geometric phase (Berry phase) is induced in
addition to the familiar dynamical phase.\cite{berry84,wilczek_book} 
A slow variation of such parameters along a closed path $C$ will return the
system to its original energy eigenstate with an additional phase factor
$\exp\{i\gamma _{n}(C)\}$. More specifically, the state acquires phases
after a period of the cycle $T$ as
\begin{equation}
|\Psi _{n}(T)\rangle = \exp\left\{-\frac{i}{\hbar} \int_0^{T}
E_{n}(t)\,dt\right\}\cdot \exp\left\{i\gamma _{n}(C)\right\}\,|\psi
_{n}\rangle.
\label{BPnon}
\end{equation}
However this equation applies only to non-degenerate states. The detailed numerical and analytical calculations of Berry phase $\gamma_n (C)$ for the Hamiltonian of a quantum dot in $2$D plane for different non-degenerate eigen states are  explained in Ref.~\onlinecite{prabhakar10}. The system
of interest here (a single spin in a $2$D electrically defined quantum
dot) is degenerate\cite{sousa03,prabhakar09} 
for which (\ref{BPnon}) is not
directly applicable. In the formulation developed by Wilczek and others\cite{wilczek_book,wilczek84} 
for degenerate cases, the geometric phase factor is
replaced by a non-Abelian unitary operator $U_{ab}$ acting on the
initial states within the subspace of degeneracy. The evolution equation
of the state is modified in the form,
\begin{equation}
|\Psi _{n, a}(t)\rangle = \exp\left\{-\frac{i}{\hbar} \int_0^{t}
E(t)\,dt\right\}\, \sum_{b} U_{ab}(t)\,|\psi _{n, b} \rangle,
\label{BPde}
\end{equation}
where $a$ and $b$ are the labels for degeneracy. The non-Abelian unitary
operator can be expressed in the form,
\begin{equation}
U_{ab}(t)=T\,\exp\left\{-\frac{i}{\hbar}\int_0^{t} {\bf A}_{ab}(t')\cdot
\dot{\bf R}\,dt'\right\},  \label{Ut}
\end{equation}
where $T$ signifies the time-ordering, and
\begin{equation}
{\bf A}_{ab}=-i\hbar \langle \psi _{n, a}\left|\nabla_{\bf R}\right|\psi
_{n,b}\rangle,  \label{Aab}
\end{equation}
${\bf R}$ and $\nabla_{{\bf R}}$ being a vector and the gradient in
parameter space, respectively.
In general, the geometric phase transformation $U_{ab}(t)$ of (\ref{Ut})
in parameter space contains non-commuting operators and time-dependent
parameters. It is possible to view the parameter-dependent evolution in
the subspace of degeneracy as a non-Abelian local gauge transformation.
Correspondingly ${\bf A}_{ab}$ in ({\ref{Aab}) may be seen as a
non-Abelian gauge connection (or the Yang-Mills fields).

Although it is not straightforward to construct the non-Abelian gauge
connection, we consider the following observation instructive for the
case where the parameter space coincides with the configuration space.
Suppose the Hamiltonian of a system is given by
\begin{equation}
H=\frac{1}{2m}({\bf P}-{\bf A})^{2} + V({\bf r}). \label{HV}
\end{equation}
The energy eigenequation $H|\psi _{n}\rangle = E_{n}|\psi _{n}\rangle $
remains invariant under the local (position-dependent) gauge transformation,
\begin{equation} |\psi _{n}\rangle \rightarrow |\psi _{n}'\rangle =
\bar{U} |\psi _{n}\rangle, ~~~~~~~{\bf A} \rightarrow  {\bf A}' =
\bar{U}{\bf A}\bar{U}^{\dagger} + i \hbar \bar{U} \nabla
\bar{U}^{\dagger}.
\end{equation}
If we choose such a gauge that the transformed vector potential
vanishes, that is, ${\bf A}'=0$, then the transformation operator is to
be of the form,
\begin{equation}
\bar{U} = \exp\left\{-\frac{i}{\hbar}\oint_c {\bf A}\cdot d{\bf r}\right\}.
\end{equation}
In other words, this transformation will ``gauge away'' the vector potential
from the Hamiltonian (\ref{HV}). Conversely, if the state with the
vanishing gauge is taken to be the initial state, the final state with
an arbitrary gauge ${\bf A}$ is obtained by the inverse gauge
transformation, $|\psi _{n}\rangle = \bar{U}^{-1}|\psi _{n}'\rangle$.
Moreover, if the inverse gauge process is time-dependent via the
variation of position, then the evolution operator is given by
\begin{equation}
U(t)=\bar{U}^{-1}(t)=T \exp\left\{\frac{i}{\hbar}\int_0^t {\bf A}\cdot \dot{\bf
r}\, dt\right\}.  \label{Ut2}
\end{equation}
This observation will be useful for our discussion on the Berry phase
associated with the spin-orbit coupling.

A matrix element of the evolution operator gives the transition
amplitude (propagator) from an initial state to the final state, which
is usually evaluated by approximation. For instance, the propagator for
the spin-orbit interaction has been calculated semiclassically in a
different context by Feynman's path integral represented in coherent
states.\cite{pletyukhov03} 

In Sec. 2, we treat the phase transformation (\ref{Ut}) as a
gauge transformation, and employ Feynman's disentangling technique,
rather than Feynman's path integral, to evaluate the time-ordered
exponential for the spin-orbit coupling Hamiltonian. Use of Feynman's
disentangling method in Popov's version\cite{popov07}  
enables us to
obtain analytical and numerical results for the spin transition
probabilities without approximation. In Sec. 3, we plot the spin-flip
probability versus the rotation angle, and compare the data for the pure
Rashba, the pure Dresselhaus, and mixed cases. Sec. 4 is devoted in
deriving analytical expressions of the non-Abelian Berry phase (the
adiabatic evolution operator as a $2 \times 2$ matrix) for the pure       
Rashba and the pure Dresselhaus coupling.
\begin{figure*}
\includegraphics[width=17cm]{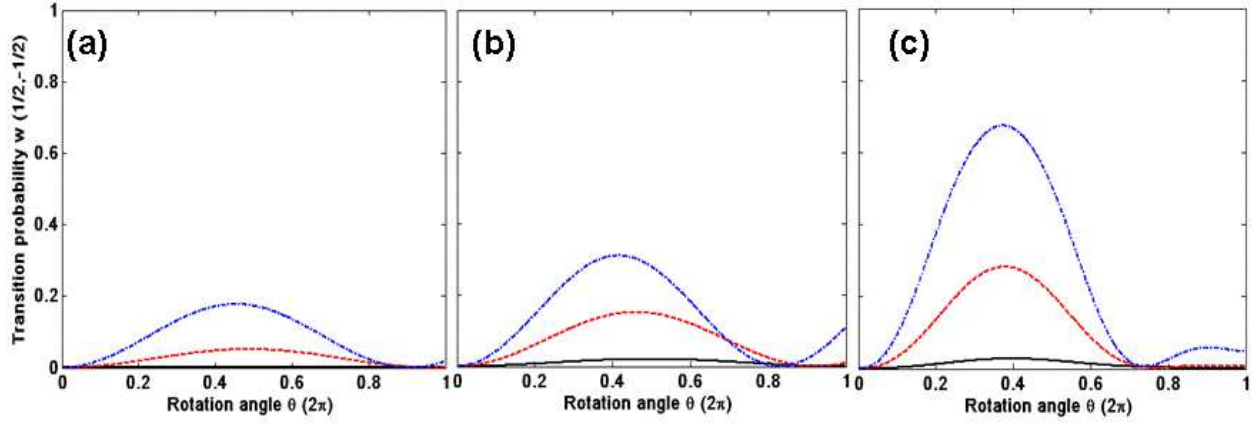}
\caption{\label{fig-a} (color online) Transition probability,
$w_{1/2, -1/2}$ vs.  $\theta$ for three cases: (a) pure
Rashba ($\beta=0$), (b) pure Dresselhaus ($\alpha=0$), and (c) mixed
(non-zero $\alpha$ and $\beta$) spin-orbit interactions.  The orbital
radius is $60$ nm. The three curves represent the following electric field
strengths: $1\times 10^5$ V/cm (solid black line), $5\times 10^5$ V/cm
(dashed red line), and $1\times 10^6$ V/cm (dotted-dashed blue line) respectively.
}
\end{figure*}
\begin{figure*}
\includegraphics[width=17cm]{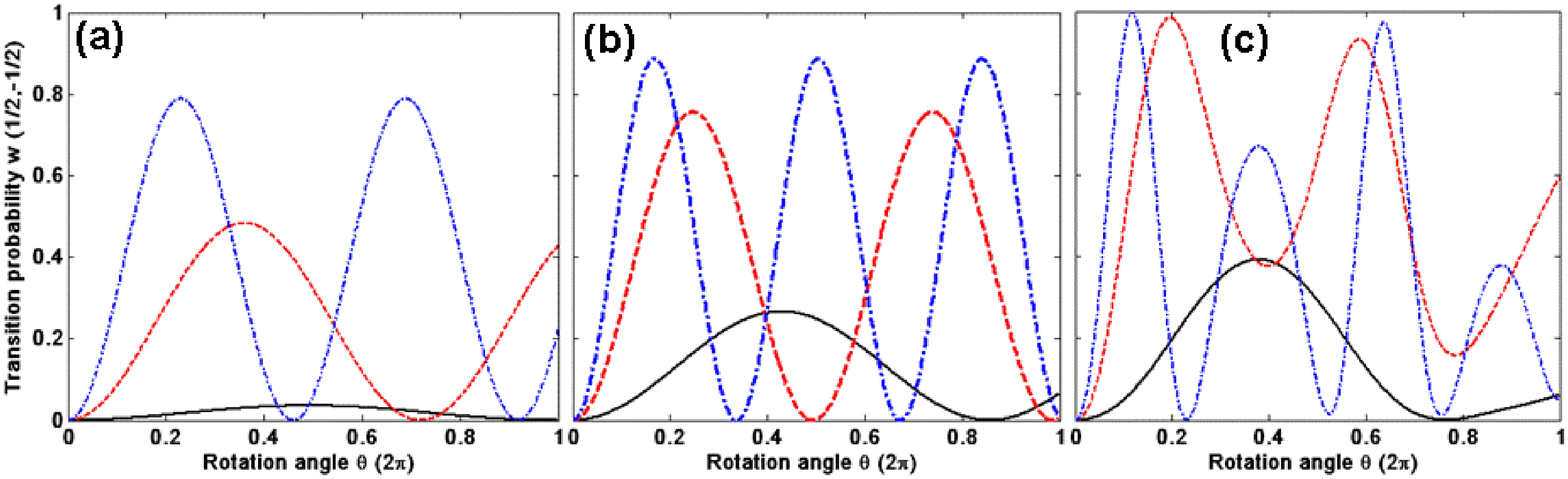}
\caption{\label{fig-b} (color online) Transition probability
$w_{1/2, -1/2}$ vs. $\theta$ for the following cases: (a) pure Rashba ($\beta=0$),
(b) pure Dresselhaus ($\alpha=0$), and (c) mixed (non-zero $\alpha$ and
$\beta$). The orbit radius is chosen to be $250$ nm and the following
values of the electric field are considered: $1\times 10^5$ V/cm (solid
black line), $5\times 10^5$ V/cm (dashed red line), and $1\times 10^6$ V/cm
(dotted-dashed blue line)}
\end{figure*}
\begin{figure*}
\includegraphics[width=17cm]{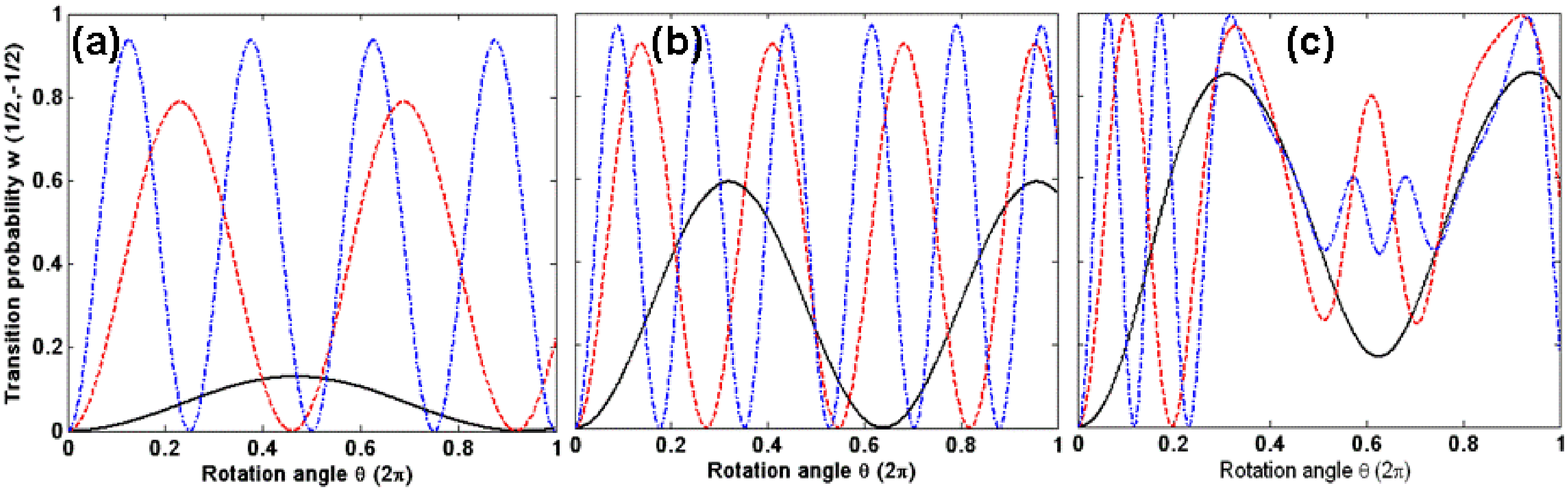}
\caption{\label{fig-c} (color online) Transition probability
$w_{1/2, -1/2}$ vs.  $\theta$ for the following cases: (a) pure Rashba
($\beta=0$), (b) pure Dresselhaus ($\alpha=0$), and (c) mixed (non-zero
$\alpha$ and $\beta$).  The orbit radius was chosen to be $500$ nm and the
following values of the electric field were chosen: $1\times 10^5$ V/cm
(solid black line), $5\times 10^5$ V/cm (dashed red line), and
$1\times 10^6$ V/cm (dotted-dashed blue line).
}
\end{figure*}
\begin{figure}
\includegraphics[width=8cm]{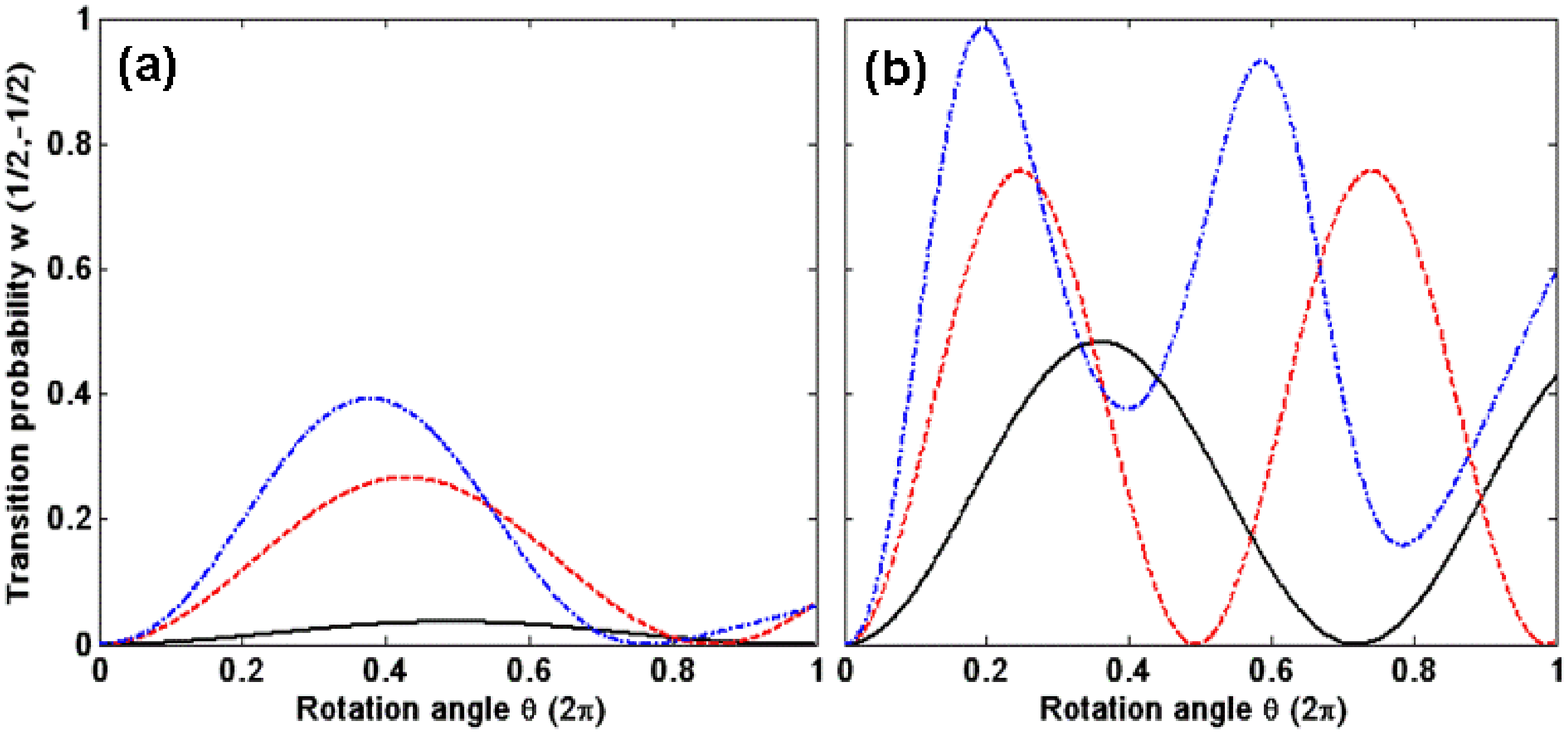}
\caption{\label{fig-d} (color online) Transition probability
$w_{1/2, -1/2}$ vs. $\theta$ for the following cases: pure Rashba
($\beta=0$: dotted-dashed blue line), pure Dresselhaus ($\alpha=0$: dashed red
line) and mixed (non-zero $\alpha$ and $\beta$: solid black line). The orbit
radius was chosen to be $250$ nm and the following values of the electric
field were chosen: (a) $E = 1\times 10^5$ V/cm and, (b) $E = 5\times10^5$ V/cm.
}
\end{figure}
\begin{figure}
\includegraphics[width=7cm]{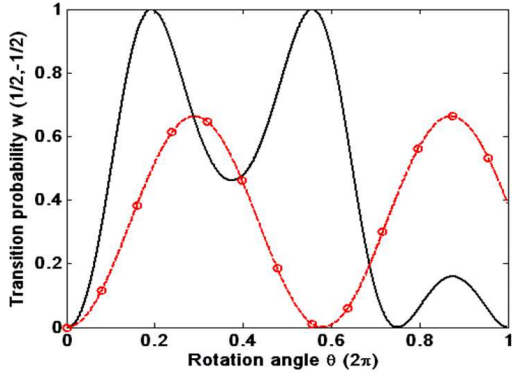}
\caption{\label{fig-e} (color online) Transition probability
$w_{1/2, -1/2}$ vs. $\theta$ for $\alpha = \beta$. Physically, this situation
occurs for electric field strength given by $E = 3.02\times 10^6$ V/cm.
Here the solid black line represents for both Rashba and Dresselhaus spin-orbit coupling effects whereas the dashed red line represents only for Dresselhaus spin-orbit coupling effect and open red circles represents only for Rashba spin-orbit coupling effect. Here we choose $60$ nm orbit radius.
}
\end{figure}
\begin{figure}
\includegraphics[width=8cm]{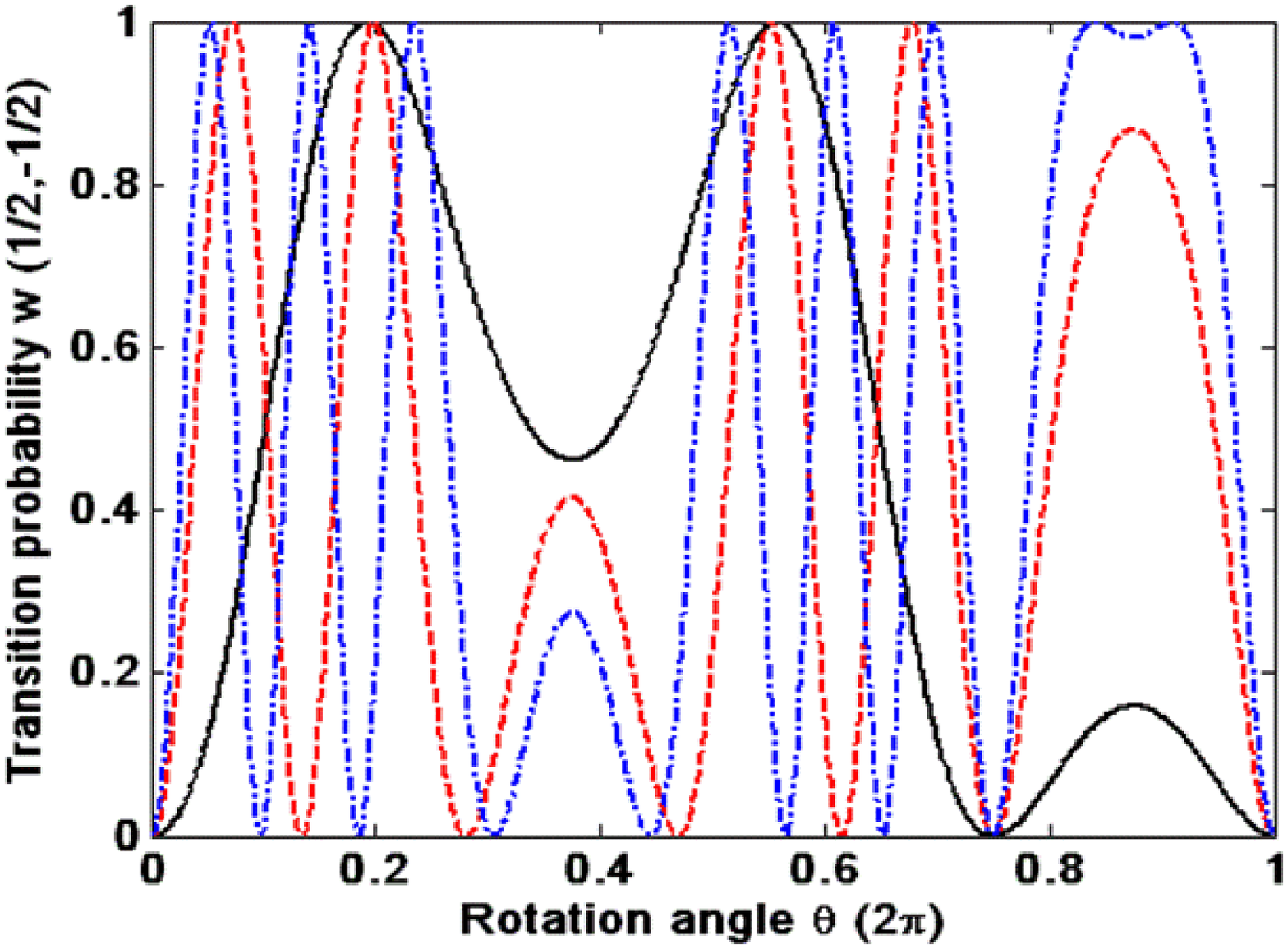}
\caption{\label{fig-f} (color online) Transition probability
$w_{1/2, -1/2}$ vs. $\theta$ for $\alpha = \beta$. Physically, this situation
occurs for electric field strength given by $E = 3.02\times 10^6$ V/cm.  The
following orbit radii were chosen: $60$ nm (solid black line), $175$ nm
(dashed red line), and, $250$ nm (dotted-dashed blue line).
}
\end{figure}


\section{SPIN TRANSITION PROBABILITIES VIA FEYNMAN DISENTANGLING METHOD}

To discuss the revolution of spin that induces a geometric phase, we consider
a GaAs quantum dot formed in the plane of a two-dimensional electron gas
($2$DEG), the center of mass of which moves adiabatically along a closed
path under the influence of applied potentials.\cite{jose06}  
The single-electron Hamiltonian in $2$DEG (in the $xy$ plane) may be written
in the form,
\begin{equation}
H=\frac{1}{2m}{\bf P}^{2} + H_{SO},   \label{Ham}
\end{equation}
where $m$ is the effective mass. The first term is the kinetic energy
in two dimensions. Evidently,  ${\bf P}^{2}=P_{x}^{2} + P_{y}^{2}$. The
second term is the spin-orbit (SO) coupling Hamiltonian in linear
approximation,
\begin{equation}
H_{SO}=2\alpha (P_{y}S_{x} - P_{x}S_{y}) - 2\beta (P_{x}S_{x} -
P_{y}S_{y}).  \label{Hso}
\end{equation}
Here ${\bf S}$ is the spin operator whose components obey
the $SU(2)$ algebra (see, e.g., Ref.~\onlinecite{inomata_book}):  
\begin{equation}
[S_{+}, S_{-}]=2S_{0}, ~~~~~ [S_{0}, S_{\pm}]=\pm S_{\pm}, \label{SU2}
\end{equation}
where $S_{\pm}=S_{x} \pm i S_{y}$ and $S_{0}=S_{z}$. The spin-orbit
Hamiltonian (\ref{Hso}) consists of the Rashba coupling whose strength
is characterized by parameter $\alpha $ and the linear Dresselhaus coupling
with $\beta $. These coupling parameters are dependent on the electric
field $E$ of the quantum well confining potential (i.e., $E=-\partial
V/\partial z$) along z-direction at the interface in a heterojunction  as
\begin{equation}
\alpha =\frac{e}{\hbar}\,a_{R}E, ~~~~~\beta =\frac{0.7794\gamma
_{c}}{\hbar}\left(\frac{2me}{\hbar^{2}} \right)^{2/3}E^{2/3},
\label{alphbeta}
\end{equation}
where $a_{R}=4.4\, \AA^{2}$ and $\gamma _{c} = 26\,eV \AA^{3}$ for
the GaAs quantum dot.~\cite{sousa03}    
The quantum well confining potential (i.e., $E=-\partial V/\partial z$) along z-direction is not symmetric in III-V type semiconductor.\cite{prabhakar09} It means, the
formation of quantum dot at the interface of III-V type semiconductor in the plane of 2DEG is asymmetric.

Now we look for the evolution operator (\ref{Ut2}) for the case of spin-orbit
coupling. It has been known that the linear spin-orbit term in
(\ref{Ham}) can be gauged away.\cite{aleiner01,chen08}  
In fact, the Hamiltonian (\ref{Ham}) may
be expressed as
\begin{equation}
H=\frac{1}{2m}\left({\bf P} - {\bf A}\right)^{2} - V_{0},
\end{equation}
where
\begin{equation}
{\bf A}=2m \left(\begin{array}{c} \alpha S_{y} + \beta S_{x} \\ \nonumber
- \alpha S_{x} - \beta S_{y}\end{array}\right) \label{A}
\end{equation}
and
\begin{equation}
V_{0}=m\hbar^{2}(\alpha ^{2} + \beta ^{2}).
\end{equation}

If the semiclassical momentum ${\bf P}=m \dot{\bf r}$ is used for the
adiabatic evolution, then the spin-orbit gauge connection is related to
the SO Hamiltonian (\ref{Hso}),
\begin{equation}
{\bf A}\cdot \dot{\bf r} = - H_{SO}.
\end{equation}
Assuming that the spin-orbit coupling is adiabatically introduced into
the initial state, we obtain via (\ref{Ut2}) the evolution operator of
the form,
\begin{equation}
U(t)=T\exp\left\{-\frac{i}{\hbar}\int_{0}^{t}H_{SO}(t')\,dt'\right\},
\label{UTt}
\end{equation}
which we shall evaluate by utilizing the Feynman disentangling method.
This form of the evolution operator is commonly employed for Berry's
phase associated with the spin-orbit interaction.\cite{jose06,prabhakar09}  

Before disentangling, we note that the SO Hamiltonian (\ref{Hso}) may
also be expressed as
\begin{equation}
H_{SO}=H_{+}S_{+} + H_{-}S_{-}
\end{equation}
with
\begin{equation}
H_{\pm}=(\alpha P_{y} - \beta P_{x}) \mp i (\beta P_{y} - \alpha
P_{x}).  \label{Hpm}
\end{equation}
Suppose the quantum dot orbits around a closed circular path of radius
$R_{0}$ in the $x-y$ plane under the influence of gate potentials, so
that ${\bf r}=R_{0}(\cos \omega t, \sin \omega t, 0)$. Then the
semiclassical momentum ${\bf P}=m\dot{\bf r}$ has components,
\begin{equation}
P_{x} = -R_{0}m\omega \,\sin \omega t, ~~P_{y}=R_{0}m\omega \,\cos
\omega t, ~~P_{z}=0.  \label{Pcom}
\end{equation}
Substitution of (\ref{Pcom}) into (\ref{Hpm}) yields
\begin{equation}
H_{\pm}=R_{0}m\omega (\alpha e^{\mp i\omega t} \mp i \beta e^{\pm \omega
t}).
\end{equation}
Since $S_{+}$ and $S_{-}$ do not commute, the evaluation of the
time-ordered exponential for the evolution operator (\ref{UTt}) is
cumbersome.

We now turn to a discussion of the Feynman disentangling technique and
its application to the present problem.
For the case where the Hamiltonian is given by
\begin{equation}
H= \alpha (t)A + \beta (t)B + \gamma (t)C  + \cdots, \label{HF}
\end{equation}
where $A$, $B$, $C$, ... are noncommuting operators, and $\alpha $,
$\beta $, $\gamma $, ... are time-dependent parameters,
Feynman~\cite{feynman51}  
devised an operator calculus by which
the time-ordered exponential can be disentangled in the form
\begin{equation}
U(t)= e^{a(t)A}e^{b(t)B}e^{c(t)C}\cdots,   \label{U2}
\end{equation}
where $a(t)$, $b(t)$, $c(t)$, ... are time-dependent
coefficients which can be determined by solving relevant differential
equations. This procedure is referred to as the Feynman
disentangling method.\cite{popov07}

Here we apply Feynman's method for disentangling the time-ordered
exponential in (\ref{UTt}) with the Hamiltonian (\ref{Hso}). First we
rewrite the Hamiltonian (\ref{Hso}) as
\begin{equation}
H_{SO}=\xi S_{+} + (H_{+}-\xi )S_{+} + H_{-}S_{-},
\end{equation}
where $\xi $ is a time-dependent function to be determined appropriately.
According to Feynman's procedure, the evolution operator may be put into
the form,
\begin{equation}
U(t)=e^{a(t)S_{+}}\,\exp\left\{\frac{1}{i\hbar}\int_{0}^{t}\,dt'\,[(H_{+}-\xi
)S'_{+} + H_{-}S'_{-}]\right\}, \label{U3}
\end{equation}
where
\begin{equation}
a(t)=\frac{1}{i\hbar}\int_{0}^{t}\xi (t')dt',
\end{equation}
\begin{equation}
S'_{+} = e^{-aS_{+}}S_{+}e^{aS_{+}}=S_{+} \label{Sp}
\end{equation}
and
\begin{equation}
S'_{-}=e^{-aS_{+}}S_{+}e^{aS_{+}}=S_{-} - 2aS_{0} - a^{2}S_{+}.
\label{Sm}
\end{equation}
Substituting (\ref{Sp}) and (\ref{Sm}) into (\ref{U3}) and choosing
$\xi(t) $ such that the coefficient of $S_{+}$ in the integrand
vanishes, we get
\begin{equation}
U(t)=e^{a(t)S_{+}}\,T\,\exp\left\{\frac{1}{i\hbar}\int_{0}^{t}\,dt'\,[-
2aH_{-}S_{0} + H_{-}S_{-}]\right\},
\end{equation}
in which the term containing $S_{+}$ is disentangled. In a similar
fashion, we disentangle the time-ordered exponential involving the
mutually non-commuting operators $S_{0}$ and $S_{-}$ by letting
\begin{eqnarray}
U(t)&=&e^{a(t)S_{+}}e^{b(t)S_{0}}\nonumber\\
&&T\,\exp\left\{\frac{1}{i\hbar}\int_{0}^{t}
\,dt'\,[(-2aH_{-} - \eta )S^{\prime \prime}_{0} + H_{-}S^{\prime
\prime}_{-}]\right\},\qquad
\end{eqnarray}
where
\begin{equation}
b(t)=\frac{1}{i\hbar}\int_{0}^{t} \eta (t')\,dt',
\end{equation}
\begin{equation}
S^{\prime \prime}_{0} =e^{-bS_{0}}S_{0}e^{bS_{0}}=S_{0}
\end{equation}
and
\begin{equation}
S^{\prime \prime}_{-}=e^{-bS_{0}}S_{-}e^{bS_{0}}=S_{-}\,e^{b}.
\end{equation}
Again choosing $\eta (t)=-2aH_{-}$, we reduce the evolution operator
(\ref{U3}) into the completely disentangled form,
\begin{equation}
U(t)=e^{a(t)S_{+}} e^{b(t)S_{0}} e^{c(t)S_{-}},  \label{Udis}
\end{equation}
where
\begin{equation}
a(t)=\frac{1}{i\hbar}\int_{0}^{t}[H_{+}(t') - a^{2}(t')H_{-}(t')]dt',
\label{inta}
\end{equation}
\begin{equation}
b(t)=-\frac{2}{i\hbar}\int_{0}^{t}a(t')H_{-}(t')\,dt' \label{intb}
\end{equation}
and
\begin{equation}
c(t)=\frac{1}{i\hbar} \int_{0}^{t}H_{-}(t')e^{b(t')}\,dt'.
\label{intc}
\end{equation}
Although the time-ordered exponential is disentangled, the evaluation of
the evolution operator remains incomplete until the coefficients $a(t)$,
$b(t)$ and $c(t)$ are determined. In general, the integral equations
(\ref{inta})-(\ref{intc}) or the equivalent differential equations
are difficult to solve. In Sec. 4, we shall determine the coefficients
and the evolution operator for the pure Rashba, and the pure Dresselhaus
coupling.

As it is seen in Appendix A, the spin transition probability depends only on
$a(t)$. Therefore the full form of the evolution operator is not needed.
To determine $a(t)$, we convert the integral equation (\ref{inta})
together with (\ref{Hpm}) into a Riccati equation of the form,
\begin{equation}
\frac{da}{dt}= -R\omega [f(t) + f^{\ast}(t)\,a^{2}(t)], \label{Ric1}
\end{equation}
where $R=mR_{0}/\hbar$,
\begin{equation}
f(t) = \beta ^{i\omega t} + i\alpha e^{-i\omega t},
\end{equation}
and
\begin{equation}
f^{\ast}(t) = \beta ^{-i\omega t} - i\alpha e^{i\omega t}.
\end{equation}
Solving (\ref{Ric1}) for $a(t)$, we can obtain the spin transition
probabilities, $w_{s,s'}$. In particular, the transition probabilities
from spin $1/2$ to $\pm 1/2$ are calculated by
\begin{equation}
w_{1/2, 1/2}=\frac{1}{1+|a|^{2}}, ~~~~~~
w_{1/2, -1/2}=\frac{|a|^{2}}{1+|a|^{2}}.  \label{prob}
\end{equation}

\section{NUMERICAL ANALYSIS}

As it is shown in Appendix B, exact solutions of the Riccati equation
(\ref{Ric1}) can be obtained only for special cases, which include those
for (i) the Rashba limit $(\beta =0)$, (ii) the Dresselhaus limit $(\alpha =0)$
and (iii) the symmetric mixture of the two couplings $(\alpha =\beta )$. The
spin-flip probabilities obtained in Appendix B for exactly solvable
cases (with $\theta =\omega t$) are:\\

\noindent(i) {\bf The Rashba limit} $(\alpha \neq 0, ~\beta =0)$:
\begin{equation}
w_{1/2, -1/2}^{R} = \frac{4R^{2}\alpha ^{2}}{1+ 4R^{2}\alpha ^{2}}\,
\sin^{2}\left(\frac{1}{2}\sqrt{1+ 4R^{2}\alpha ^{2}}\,\theta \right);
\label{pRash}
\end{equation}

\noindent(ii) {\bf The Dresselhaus limit} $(\alpha = 0, ~\beta \neq 0)$:
\begin{equation}
w_{1/2, -1/2}^{D} = \frac{4R^2\beta^2}{1+4R^2\beta^2}\sin^2\left\{ \frac{1}{2}\sqrt{1+4R^2\beta^2}\,\,\,\theta\right\};
\label{pDres}
\end{equation}
\noindent(iii) {\bf The symmetric Rashba-Dresselhaus limit} $(\alpha = \beta
\neq 0)$:
\begin{equation}
w_{1/2, -1/2}^{sym} = \sin^{2}\left\{\sqrt{2}\alpha R
(\sin \theta  - \cos \theta  + 1)\right\}. \label{symm}
\end{equation}

For an arbitrarily mixed Rashba-Dresselhaus coupling (mixed R-D),
the Riccati equation (\ref{Ric1}) is not exactly solvable.
Therefore numerical analysis is needed. In the below we treat
the mixed R-D coupling $(\alpha \neq \beta )$ and the symmetric R=D
coupling $(\alpha = \beta )$ separately.\\

\noindent{\bf Comparison of the Rashba coupling, the Dresselhaus
coupling and the mixed R-D coupling}:  Figs.~\ref{fig-a},~\ref{fig-b} and ~\ref{fig-c} 
plot the spin-flip probability $w_{1/2, -1/2}$ versus the rotation angle $\theta
=\omega t$ in the unit of $2\pi $ for the orbit radius $R_0$=$60$ nm, $250$
nm, and $500$ nm, respectively. The plots of (a), (b) and (c) in these
figures correspond to (a) the pure Rashba case $(\beta =0)$, (b) the
pure Dresselhaus case $(\alpha =0)$ and (c) the mixed R-D case $(\alpha
\neq 0, \beta \neq 0)$, respectively. The three different values of the
electric field $E= 1 \times 10^{5}$ V/cm, $5 \times 10^{5}$ V/cm, and $1
\times 10^{6}$ V/cm, are chosen for the curves in each figure, solid
black, dashed red, and dotted-dashed blue, respectively. The symmetric case
(R=D) will be examined separately with Figs.~\ref{fig-e} and ~\ref{fig-f}.

The curves for (a) the pure Rashba case and (b) the pure Dresselhaus case
are obtained from the exact results (\ref{pRash}) and (\ref{pDres}). As it
is obvious from these equations, the spin-flip probability increases as
the electric field increases via the coupling parameter but remains to be
less than unity. Another observation we can make from these plots is
that the periods of spin-flip for the pure Rashba coupling and the pure
Dresselhaus coupling are different. This is also expected from the
analytical results (\ref{pRash}) and (\ref{pDres}).

The curves in Figs.~\ref{fig-a}(c), ~\ref{fig-b}(c) and ~\ref{fig-c}(c) 
show the spin-flip probability for (c) the mixed R-D case where both $\alpha $ and $\beta $
are not zero and not equal. Note that they are {\it not} the results
from the exact formula (\ref{symm}) for the symmetric R-D
coupling. Since the Riccati equation (\ref{Ric1}) for
arbitrary non-zero $\alpha $ and $\beta $ is not solvable, we carry out
numerical simulations by using numerical solutions of (\ref{Ric1}) in
(\ref{prob}). The spin-flip probability for the mixed case is generally
larger than the pure cases. Furthermore, it does not reach unity if
$\alpha \neq \beta $. In other words, the complete spin-flip is not
likely to occur during the entire period of the adiabatic motion along
the closed orbit. In the vicinity of the symmetry point $(\alpha
=\beta )$, the transition probability becomes very close to unity at
certain angles.

Fig.~\ref{fig-d} 
gives a further comparison study of the transition probability
for the pure Rashba, the pure Dresselhaus, and the mixed case. In Fig.~\ref{fig-d}(a), when the electric field is weak, the curve for the mixed case
appears to be a superposition of those for the two pure cases. As the electric
field increases, the superposition effect becomes obscure as is seen in
Fig.~\ref{fig-d}(b). As the Riccati equation is nonlinear in nature, there is no
reason to expect that the mixed case is a superposition of the two pure
cases. It is interesting to observe that the mixed case has a better
chance to achieve the spin-flip than the pure cases during the period of
evolution.\\

\noindent{\bf Analysis of the symmetric R-D coupling}:-
The symmetric mixture of the Rashba and Dresselhaus couplings has been
discussed in connection with the persistent spin helix.~\cite{bernevig06,liu05}  
Bernevig {\it et al.}~\cite{bernevig06} 
found an exact $SU(2)$ symmetry in the
symmetric mixture and predicted the persistent spin helix which is
a helical spin density wave with conserved amplitude and phase.
Recently spin life time enhancement of two orders of magnitude near the
symmetry point $(\alpha =\beta)$ has been reported experimentally.~\cite{koralek09}  

The coupling parameters $\alpha $ and $\beta $ of the Rashba and
Dresselhaus interactions are given by (\ref{alphbeta}) for the GaAs
quantum dot. The two parameters become equal at $E=3.02 \times
10^{6}$ V/cm. For the situation in which the two couplings have equal
strength (i.e., $\alpha =\beta $), the Riccati equation (\ref{Ric1}) is
exactly solved and the corresponding transition probability is given by
(\ref{symm}). In Fig. 5, the spin-flip probability versus the angle of
rotation along the orbit of radius 60 nm is plotted at $E=3.02 \times
10^{6}$ V/cm for the pure Rashba case (open red circles), the pure
Dresselhaus case (dashed red line), and the symmetric case (solid black
line). We see that the symmetric Rashba-Dresselhaus spin-orbit coupling definitely achieves a
spin-flip during the adiabatic process whereas the two pure
cases have less chances. Fig.~\ref{fig-f} 
plots the transition probability of the
symmetric R-D case for three different radii of the orbit of the quantum
dot: 60 nm (solid black line), 175 nm (dashed red line) and 250 nm
(dotted-dashed blue line). It shows that the chance of being in the
spin-flip state is enhanced by increasing the orbit radius.

It is important to notice that the complete spin flip takes place only
in the symmetric R-D coupling. This may be an indication of the persistent
spin helix. Although the assumed orbit of motion is circular, we can
regard the motion for a small angle of rotation as linear. Let $\theta
=\varepsilon \approx 0$ or $\theta = 3\pi /2 -\varepsilon $. If
$\varepsilon $ is small, then $\sin \theta  - \cos \theta +1 \approx
\varepsilon $, and the exact formula (\ref{symm}) may be approximated by
\begin{equation}
w_{1/2, -1/2}^{sym}=\sin^{2}\left\{\sqrt{2}\alpha R \varepsilon \right\}.
\end{equation}
As $\varepsilon $ varies from 0 to $\pi /(2\sqrt{2}\alpha R)$, the
spin-flip probability moves from zero to unity, that is, the spin
completes a full precession. For instance, if $R=60$ nm, the range $0
\leq \sqrt{2}\alpha R \varepsilon \leq \pi /2 $ corresponds to the
portion of the solid black curve for $ 0 \leq \theta /2\pi < 0.2$ in
Fig.~\ref{fig-e}. 
Let $\varepsilon _{s}=\pi /(2\sqrt{2}\alpha R)$. Then
the $R\varepsilon _{s}$ is the distance the electron progresses
while the spin precesses by $2\pi $. Therefore, we may be able to
identify this distance with the spin diffusion length $L_{s}$ as
\begin{equation}
L_{s}=R\varepsilon _{0}/\pi =\frac{1}{2\sqrt{2}\alpha}.
\end{equation}

\section{ANALYTICAL EXPRESSION FOR THE NON-ABELIAN BERRY PHASE}

Applying the Feynman disentangling method, we have been able to reduce
the time-ordered evolution operator (\ref{UTt}) to the disentangled form
(\ref{Udis}) with the time-dependent scalar functions $a(t)$, $b(t)$
and $c(t)$ obeying the integral equations (\ref{inta})-(\ref{intc}).
It is sometimes convenient to express the evolution operator
as a $2\times 2$ matrix in the spin representation of $SU(2)$.
Evidently the $SU(2)$ algebra (\ref{SU2}) is satisfied by
\begin{equation}
S_{+}=\left(\begin{array}{cc}0 & 1\\ \nonumber 0 & 0 \end{array}\right),
~~
S_{0}=\frac{1}{2} \left(\begin{array}{cc}1 & 0\\ \nonumber 0 & -1
\end{array}\right),~~
S_{-}=\left(\begin{array}{cc}0 & 0\\ \nonumber 1
& 0 \end{array}\right).
\end{equation}
Using the properties $S_{\pm}^{2}=0$ and $S_{0}^{2}=1/4$, we can
write (\ref{Udis}) as
\begin{equation}
U(t)=\left(\begin{array}{cc}1 & a\\ \nonumber 0
& 1 \end{array}\right)\left(\begin{array}{cc}e^{b/2} & 0\\ \nonumber 0
& e^{-b/2} \end{array}\right)
\left(\begin{array}{cc}1 & 0\\ \nonumber c
& 1 \end{array}\right),
\end{equation}
from which immediately follows that
\begin{equation}
U(t)=\left(\begin{array}{cc}e^{b/2} + ac\,e^{-b/2} & ~ae^{-b/2}\\ \nonumber
c\,e^{-b/2} & e^{-b/2} \end{array}\right). \label{matri}
\end{equation}
This is the desired matrix representation for the Berry phase, and is
used for calculating the spin-flip probabilities in Appendix A.

The expressions (\ref{Udis}) and (\ref{matri}) remain formal until the
time-dependent functions $a(t)$, $b(t)$ and $c(t)$ are specified.
Eq. (\ref{inta}) for $a(t)$ is equivalent to a Riccati equation without
whose solution, (\ref{intb}) and (\ref{intc}) cannot be solved for $b(t)$
and $c(t)$. In Appendix B, we show that the Riccati equation can be
solved exactly if the function $h(t)$ defined by
\begin{equation}
h(t)=\frac{\beta ^{2}-\alpha ^{2}}{2R(\alpha ^{2} + \beta
^{2})^{3/2}}\left[1 + \frac{2\alpha \beta }{\alpha ^{2} + \beta
^{2}}\sin(2\omega t)\right]^{-3/2} \label{ht}
\end{equation}
becomes time-independent $(h(t)=h_{0})$. The last restriction (\ref{ht})
is fulfilled only when one of the following conditions is met: $\alpha =0$,
$\beta =0$ or $\alpha =\beta $. This implies that the function $a(t)$
can be determined only for the pure Dresselhaus coupling, the pure
Rashba coupling and the symmetric Rashba-Dresselhaus coupling. The
result we find for $a(t)$ is
\begin{equation}
a(t)=\frac{if}{|f|}\frac{e^{i\phi(t)} -1}{n_{1}e^{i\phi(t)}-n_{2}},
\end{equation}
where
\begin{equation}
f(t)=\beta e^{i\omega t} + i\alpha e^{-i\omega t},
\end{equation}
\begin{equation}
\phi (t)=R\omega (n_{1}-n_{2})\int_{0}^{t}|f(t')|dt'.
\end{equation}
Here
\begin{equation}
n_{1}, n_{2}=h_{0} \pm \sqrt{h_{0}^{2} +1}.
\end{equation}
For convenience, we choose $n_{1} > n_{2}$.
A closed form expression for $\phi (t)$ is given in (B12).

In calculating the spin transition probability, all we need is $a(t)$.
However, for completing the evolution operator we have also to determine
other functions $b(t)$ and $c(t)$ by solving (\ref{intb}) and
(\ref{intc}) for the already determined function $a(t)$. As has been
mentioned above, the Riccati equation can be solved exactly for the pure
Rashba coupling, the pure Dresselhaus coupling and the symmetric
Rashba-Dresselhaus coupling. In the two pure couplings, the phase
function $\phi (t)$ can be expressed in the form,
\begin{equation}
\phi (t)=\varphi t, \label{lamb}
\end{equation}
where $\varphi =\sqrt{1 + 4\alpha ^{2}R^{2}}\,\omega $ for the Rashba
coupling and $\varphi =\sqrt{1 + 4\beta ^{2}R^{2}}\,\omega $ for the
Dresselhaus coupling. For the symmetric R-D coupling, it cannot be
simplified in the form of (\ref{lamb}). Therefore, it is difficult to
carry out integration in (\ref{intb}) and (\ref{intc}). This means that
we have analytical expressions of the adiabatic evolution operator
(\ref{matri}) only for the pure Rashba and the pure Dresselhaus cases.
For the symmetric R-D coupling, even though we have no analytical expressions for
$b(t)$ and $c(t)$, we can calculate the spin-flip probability since
$a(t)$ is found in closed form.

In what follows we provide the results of integration for the pure
Rashba coupling and the pure Dresselhaus coupling.\\

\noindent(i) {\bf The pure Rashba coupling} $(\alpha \neq 0, ~\beta =0)$:

In this case, (B2), (B4), (B5) and (B11) yield,
\[
if/|f|=-e^{i\omega t}, ~~~~~h_{0}=-\frac{1}{2\alpha R}
\]
and
\[
\phi (t)=\varphi t, ~~~~~~\varphi =\omega \sqrt{1 + 4\alpha ^{2}R^{2}}.
\]
Upon substitution of these results into (B13) we arrive at
\begin{equation}
a(t)=-e^{i\omega t}\frac{e^{i\varphi t} - 1}{n_{1}e^{\varphi
t} - n_{2}}, \label{arash}
\end{equation}
where
\[
n_{1}, ~n_{2}=-\frac{1}{2\alpha R} \pm \frac{1}{2\alpha R}\sqrt{1 +
4\alpha ^{2}R^{2}}.
\]
From Eq.(\ref{intb}), using
\[
H_{-}=\alpha R \hbar \omega e^{i\omega t},
\]
together with (\ref{arash}), we obtain
\begin{equation}
e^{b(t)}=\frac{(n_{1}-n_{2})^{2}\,e^{i(\varphi -\omega
)t}}{(n_{1}e^{i\varphi t}-n_{2})^{2}},
\end{equation}
and
\begin{equation}
c(t)=\frac{1-e^{i\varphi t}}{n_{1}e^{i\varphi t} - n_{2}}.
\end{equation}
~\\

\noindent(ii) {\bf The pure Dresselhaus coupling} $(\alpha =0, ~\beta \neq 0)$:

In this case, we have
\[
if/|f|=ie^{i\omega t}, ~~~~~h_{0}=\frac{1}{2\beta R}
\]
and
\[
\phi (t)=\varphi t, ~~~~~~\varphi =\omega \sqrt{1 + 4\beta ^{2}R^{2}}.
\]
Hence we get
\begin{equation}
a(t)=ie^{i\omega t}\frac{e^{i\varphi t} - 1}{n_{1}e^{\varphi
t} - n_{2}}, \label{adress}
\end{equation}
where
\[
n_{1}, ~n_{2}=-\frac{1}{2\beta R} \pm \frac{1}{2\beta R}\sqrt{1 +
4\beta ^{2}R^{2}}.
\]
Use of
\[
H_{-}=i\beta R \hbar \omega e^{-i\omega t},
\]
and (\ref{adress}) leads to
\begin{equation}
e^{b(t)}=\frac{(n_{1}-n_{2})^{2}\,e^{i(\varphi + \omega
)t}}{(n_{1}e^{i\varphi t}-n_{2})^{2}},
\end{equation}
~\\
~\\
and
\begin{equation}
c(t)=-i\frac{1-e^{i\varphi t}}{n_{1}e^{i\varphi t} - n_{2}}.
\end{equation}

\section{CONCLUSION}

In the present paper we have considered spin manipulation via the
non-Abelian Berry phase induced by an adiabatic transport of a single
spin along a circular path in the 2D plane in the presence of the Rashba
and Dresselhaus spin-orbit couplings. We have adopted the Feynman
disentangling technique to calculate the spin-flip probability. We have
shown that the problem can be solved exactly in three cases: (i) the
pure Rashba coupling, (ii) the pure Dresselhaus coupling, and (iii) the
symmetric combination of Rashba and Dresselhaus couplings. For
an arbitrary combination of the two couplings, we have carried out
numerical simulations. We have plotted the spin-flip probability versus
the angle of the adiabatic rotation with various values of the electric
field and the radius of the circular path in the 2D plane. We have
observed that a complete spin-flip (a complete spin precession) occurs
only when the strength of the two couplings becomes equal. The relation
between the complete spin precession and the persistent spin helix will
be discussed in detail elsewhere. We have also obtained analytical
expressions of the non-Abelian Berry phase for the pure Rashba case and
the pure Dresselhaus case.\\[2mm]

\appendix

\section{The spin transition probabilities}

Following Popov's procedure,\cite{popov58,popov07}  
we show that the
spin-flip probability can be expressed in the form of (\ref{prob}).
Since the time-evolution of the spin state can be achieved by a time-dependent rotation, the transition amplitude for spin  $\sigma $ to
$\sigma '$ is given by
\begin{equation}
\langle \sigma |U(t)| \sigma
'\rangle = D^{s}_{\sigma ,\sigma ^{\prime}}(\varphi , \vartheta, \phi
)=\exp\left[-i(\sigma \varphi + \sigma '\phi )\right]d_{\sigma \sigma
'}(\vartheta ).
\end{equation}
Here $\varphi $, $\vartheta$, $\phi $ are the time-dependent Eulerian
angles, $ D^{s}_{\sigma ,\sigma ^{\prime}}(\varphi , \vartheta, \phi )$
are the elements of the Wigner $D$-matrix being the irreducible unitary
representations of $SU(2)$ group, and $d_{\sigma \sigma '}(\vartheta )$
is Wigner's $d$-function.

The corresponding transition probability along the $z$-axis is
\begin{equation}
w_{\sigma \sigma '}=|d_{\sigma \sigma '}^{s}(\vartheta )(t)|^{2}.
\end{equation}
In particular, the transition probability from spin $1/2$ to $\pm 1/2$ is
\begin{equation}
w_{1/2,1/2}=\cos^{2}\left(\frac{\vartheta(t)}{2}\right), \label{A3}
\end{equation}
and
\begin{equation}
w_{1/2,-1/2}=\sin^{2}\left(\frac{\vartheta(t)}{2}\right), \label{A4}
\end{equation}
because
\begin{equation}
d_{1/2, 1/2}^{1/2}(\vartheta )=\cos\frac{\vartheta }{2}, ~~~~~~~~
d_{1/2, -1/2}^{1/2}(\vartheta )=i\sin\frac{\vartheta }{2}.
\end{equation}

For spin $s=1/2$, the rotation matrix is given in the
standard form,\cite{popov07,inomata_book}  
\begin{equation}
{\bf D}(\varphi , \vartheta, \psi )=\left(\begin{array}{cc}
\tilde{\alpha }  & -\tilde{\beta }^{\ast}\\ \tilde{\beta }  & \tilde{\alpha
}^{\ast}
\end{array}\right),
\end{equation}
where
\begin{equation}
\tilde{\alpha}=\cos\frac{\vartheta }{2}\,\exp\left[i\frac{\psi
+\varphi}{2}\right],
\tilde{\beta }=i \sin \frac{\vartheta }{2}\,\exp\left[i\frac{\psi
-\varphi}{2}\right].
\end{equation}

Comparison of the evolution operator for the spin 1/2 transition
expressed in the matrix form,
\begin{equation}
{\bf U}=\left(\begin{array}{cc} e^{b/2}+ace^{-b/2} & ae^{-b/2}\\
ce^{-b/2} & e^{-b/2}\end{array}\right),
\end{equation}
and the rotation matrix yields
\begin{equation}
|a|^{2} = \tan^{2}\frac{\vartheta }{2}.
\end{equation}
Again comparing this result with (A3) and (A4), we arrive at
\begin{equation}
w_{1/2, 1/2} = \frac{1}{1+|a|^{2}}, ~~~~~~~
w_{1/2, -1/2} = \frac{|a|^{2}}{1+|a|^{2}}.
\end{equation}
Note that $w_{1/2, 1/2} + w_{1/2, -1/2}=1$.\\[2mm]

\section{Special solutions of the $\alpha -\beta $ Riccati equation}

Here we wish to solve under a special condition the Riccati equation
(\ref{Ric1}):
\begin{equation}
\frac{da(t)}{dt}=-R\omega \{f(t) + f^{\ast}(t
)\,a^{2}(t)\},  \label{B1}
\end{equation}
where
\begin{equation}
f(t)=\beta\,e^{i\omega t} + i\alpha\,e^{-i\omega t}, ~~~~~
f^{\ast}(t)=\beta\,e^{-i\omega t} - i\alpha\,e^{i\omega t}.
  \label{B2}
\end{equation}
This equation contains the Rashba limit $(\alpha \neq 0, ~\beta =0)$,
and the Dresselhaus limit $(\alpha =0, ~\beta \neq 0)$, both of which
have exact solutions.\\

First we let $a(t)=g(t)\,X(t)$ in (B1). If we further let
$g(t)=-if/|f|$, then we see that $X(t)$ obeys
\begin{equation}
\frac{dX}{dt} = iR\omega |f(t)|
\left\{X^{2} - 2h(t)X -1\right\},     \label{B3}
\end{equation}
where
\begin{equation}
|f(t)|=\left[\alpha ^{2} + \beta ^{2} + 2\alpha \beta \sin(2\omega t)
\right]^{1/2}     \label{B4}
\end{equation}
and
\begin{equation}
h(t)=\frac{\beta ^{2}-\alpha ^{2}}{2R(\alpha ^{2} + \beta
^{2})^{3/2}}\left[1 + \frac{2\alpha \beta }{\alpha ^{2} + \beta
^{2}}\sin(2\omega t)\right]^{-3/2}.    \label{B5}
\end{equation}

Now we consider a special case where $h(t)$ is a constant, say,
$h_{0}$. In this case, (B3) can be expressed as
\begin{equation}
\frac{dX}{(X-n_{1})(X-n_{2})} = iR\omega \,|f(t)| \,dt,  \label{B7}
\end{equation}
where $n_{1}$ and $n_{2}$ are roots of
\begin{equation}
X^{2}-2h_{0}X-1=0,
\end{equation}
that is,
\begin{equation}
n_{1}, ~n_{2} = h_{0} \pm \sqrt{h_{0}^{2} + 1}.
\label{B8}
\end{equation}
Note that
\begin{equation}
n_{1}n_{2}=-1, ~~~~n_{1}+n_{2}=2h_{0}, ~~~~n_{1}-n_{2}=2\sqrt{h_{0}^{2}+1}.
\end{equation}
Upon integration, we obtain with the condition $X(0)=0$,
\begin{equation}
X(t)=-\frac{1-e^{i\phi (t)}}{n_{2}-n_{1}e^{i\phi (t)}}.
\label{B10}
\end{equation}
The phase function $\phi (t)$ is
\begin{equation}
\phi(t)=R\omega (n_{1}-n_{2})\int_{0}^{t}|f(\tau )|\,d\tau
\label{B11}
\end{equation}
which can be expressed in closed form,
\begin{eqnarray}
\phi (t)&=&2R\omega \sqrt{h_{0}^{2}+1}(\alpha +\beta )\nonumber\\
&&\left\{E\left(\omega t -
\frac{\pi }{4}, \frac{2\sqrt{\alpha \beta }}{\alpha +\beta }
\right) - E\left(-
\frac{\pi }{4}, \frac{2\sqrt{\alpha \beta }}{\alpha +\beta }
\right)\right\},\qquad\,\,\,
\end{eqnarray}
where $E(\varphi , k)$ is the elliptic function of the second kind
defined by
\[
E(\varphi , k)=\int_{0}^{\varphi } \sqrt{ 1 - k^{2}\sin^{2}\theta }
\,d\theta.
\]

Consequently, for the case where $h(t)=h_{0}$, the starting Riccati
equation (B1) is exactly solved, the result being of the
form,
\begin{equation}
a(t)=\frac{if}{|f|}\frac{e^{i\phi}-1}{n_{1}e^{i\phi} -
n_{2}}. \label{B13}
\end{equation}
Since $\phi (0)=0$, it is evident that $a(0)=0$.
Using (B8) in (B13), we obtain
\begin{equation}
|a(t)|^{2} =\frac{\sin^{2}(\phi /2)}{h_{0}^{2}+1 - \sin^{2}(\phi
/2)}.
\end{equation}
The transition
probabilities from spin $1/2$ to $\pm 1/2$ are given by
\begin{equation}
w_{1/2, 1/2}=1 - \frac{1}{h_{0}^{2}+1}\sin^{2}\left(\frac{\phi
(t)}{2}\right)
\end{equation}
and
\begin{equation}
w_{1/2,-1/2}=\frac{1}{h_{0}^{2}+1}\sin^{2}\left(\frac{\phi
(t)}{2}\right),
\end{equation}
which are characterized only by the constant $h_{0}$ and the phase
function $\phi (t)$.

Although the above results are exact under the assumption that
$h(t)=h_{0}$ is a constant, they are approximate results
when $h(t)\approx h_{0}$.

Finally, specifying the values of $h_{0}$ and $\phi (t)$, we shall obtain the
exact results for the Rashba, the Dresselhaus and the symmetric cases.\\

\noindent(i) {\bf The Rashba limit} $(\alpha \neq 0, ~\beta =0)$: In this
case, from (B5) follows
\begin{equation}
h_{0}=-\frac{1}{2\alpha R\omega }.
\end{equation}
Furthermore the right-hand side of (B11) can be easily integrated, so
that
\begin{equation}
\phi (t)=\sqrt{1+ 4 \alpha ^{2}R^{2}}\,\omega t.
\end{equation}
Thus the spin flip probability is obtained in the form,
\begin{equation}
w_{1/2, -1/2}^{R} = \frac{4\alpha ^{2}R^{2}}{1+ 4\alpha ^{2}R^{2}}
\,\sin^{2}\left\{\frac{1}{2}\sqrt{1+ 4\alpha ^{2}R^{2}}\,\theta
\right\},
\end{equation}
where $\theta =\omega t$.\\

\noindent(ii) {\bf The Dresselhaus limit} $(\alpha = 0, ~\beta \neq 0)$: In this
case, (B5) leads to
\begin{equation}
h_{0}=\frac{1}{2\beta R\omega }.
\end{equation}
The integral of (B11) yields
\begin{equation}
\phi (t)=\sqrt{1+ 4\beta ^{2}R^{2}}\,\omega t.
\end{equation}
Hence the spin-flip probability is
\begin{equation}
w_{1/2, -1/2}^{D} = \frac{4\beta ^{2}R^{2}}{1+ 4\beta ^{2}R^{2}}
\,\sin^{2}\left\{\frac{1}{2}\sqrt{1+ 4\beta ^{2}R^{2}} \theta
\right\}.
\end{equation}

\noindent(iii) {\bf The symmetric case} $(\alpha = \beta \neq 0)$: In this
particular case,
\begin{equation}
h_{0}=0.
\end{equation}
The phase factor becomes
\begin{equation}
\phi (t)=2\sqrt{2}\alpha R\left[\sin(\omega t) - \cos(\omega t)
+ 1\right].
\end{equation}
The corresponding spin-flip probability as a function of $\theta =\omega t$ is
\begin{equation}
w_{1/2, -1/2}^{sym} = \sin^{2}\left\{\sqrt{2}\alpha R
[\sin \theta  - \cos \theta  + 1]\right\}.
\end{equation}
~\\

\begin{acknowledgments}
This work was supported by the NRI INDEX center, USA,  NSERC and CRC program, Canada.
\end{acknowledgments}

\end{document}